\documentclass[aps,prc,preprint,showpacs,nofootinbib]{revtex4-1}
\usepackage{graphics}
\usepackage{epsfig}

\newcommand\ba{\begin{eqnarray}}
\newcommand\ea{\end{eqnarray}}
\newcommand\be{\begin{equation}}
\newcommand\ee{\end{equation}}

\begin{document}
\title{ Multiplicity description by gluon model}
\author{E.~S.~Kokoulina}
\affiliation{\it JINR, VBLHE, Dubna, Moscow region, Russian
Federation, 141980}
\email{kokoulin@sunse.jinr.ru}
\date{\today}
\begin{abstract}
Study of high multiplicity events in proton-proton interactions is carried out at the U-70~accelerator (IHEP, Protvino). These events are extremely rare. Usually, Monte Carlo codes underestimate topological cross sections in this region. The gluon dominance model (GDM) was offered to describe them. It is based on QCD and a phenomenological scheme of a hadronization stage. This model indicates a recombination mechanism of hadronization and a gluon fission. Future program of the SVD Collaboration is aimed at studying  a long-standing puzzle of excess soft photon yield and its connection with high multiplicity at the U-70 and Nuclotron facility at JINR, Dubna.

\end{abstract}
\maketitle
\section{Introduction}

Multi-particle production remains an actual theme of modern high energy physics. The SVD Collaboration studies proton-proton interactions at the U-70 accelerator of IHEP, Protvino \cite{SVD} (experiment E-190) with a high number of secondary particles (mainly, pions)
\begin{equation}
p+p \rightarrow 2N+N_\pi \pi,
\end{equation}
where $N$ is a nucleon and $N_\pi $ is a number of pions. The energy of the proton beam is equal to 50~GeV, the average charged multiplicity $\overline N_{ch}$ at this energy is equal to 5.45. Pions are hadrons copiously produced at this energy. We study the high charged ($N_{ch}$) and high total ($N_{tot}$) multiplicities that are larger than the average ones. The kinematical limit is defined from a condition of all kinetic energy transformation of colliding protons into the mass of pions
\begin{equation}
N_{thresh}\simeq (\sqrt s - 2 m_p)/m_{\pi }.
\end{equation}

At 50 GeV/c the kinematical limit is equal to approximately 59 pions. The experiment is carried out at the SVD (Spectrometer with Vertex Detector) setup located at the U-70 accelerator of IHEP, Protvino \cite{SVD,SVD1}. Its main elements are a hydrogen target, a silicon vertex detector, a drift tube tracker, a magnetic spectrometer (proportional chambers and a large magnet) and an electromagnetic calorimeter \cite{SVD2}. The setup registers charged particles and photons. The high charged multiplicity events are extremely rare. The scintillator hodoscope (a high multiplicity trigger) was manufactured to suppress registration of small multiplicity events and to register events with multiplicity higher than a given level \cite{Trig}.

It is known that Monte-Carlo event generators are often mistaken in their predictions of topological cross sections in the high multiplicity region \cite{Phenom}. They usually underestimate them. The existent models give diverse predictions of high multiplicity behaviour \cite{Phenom1}. So the experimental and following theoretical studies are necessary. The SVD Collaboration has advanced in measuring topological cross sections of \emph{pp} interactions \cite{SVD,SVD1} with high charged multiplicity. Previous Mirabelle data  \cite{Mirab} were renovated for $N_{ch}$ from 10 up to 16 and we have added 4 new points from 18 up to 24. The topological cross section at the last observed point, $N_{ch}$~=~24, is three orders of magnitude lower than it was obtained by the Mirabelle Collaboration  at $N_{ch}$ = 16.

Neutral particles, photons, are registered by the electromagnetic calorimeter (ECal). Owing to its restricted acceptance it is impossible to restore all neutral pions directly. That is why an original algorithm was developed to define a number of events with a certain multiplicity of $\pi  ^0$-mesons \cite{Fluct,Meson,FluRiad,ICHEP12,Bald12} and a number of events with the total multiplicity, $N_{tot}=N_{ch}+N_{0}$, where $N_0$ is the multiplicity of neutral pions.

Some collective phenomena are predicted in this region. the SVD Collaboration is aimed to search for Bose-Einstein condensation (BEC) of pions \cite{Land,Goren1,Goren}, the peak structure at the angular distributions stipulated for the Cherenkov radiation or the shock wave formation \cite{Ulery}, and the study of the anomalous yield of soft photons ($p_T < $ 100 MeV) \cite{Chlia} in high multiplicity events. 

The noticeable growth of a scaled variance (ratio of variance to average multiplicity) was revealed  in proton-proton interactions in the region of high total multiplicity \cite{Fluct,Meson,FluRiad,ICHEP12,Bald12}. In accordance with the Begun-Gorenstein model \cite{Land,Goren1,Goren} of an ideal hadron gas,  it can be one of the indications of the pion BEC formation. The two-humped structure in angular distribution is preliminary observed in events with multiplicity higher than the average one \cite{Braz,cher}. For prediction of behaviour of the topological cross sections  in the high multiplicity region the  gluon dominance model (GDM) was offered \cite{GDM1,GDM2}. Studying the soft photon yield puzzle is our following investigation at the U-70 and Nuclotron (JINR, Dubna) \cite{SP,Barsh,SP1}.  

\begin{figure}[t]
\centerline{
\includegraphics[width=14cm]{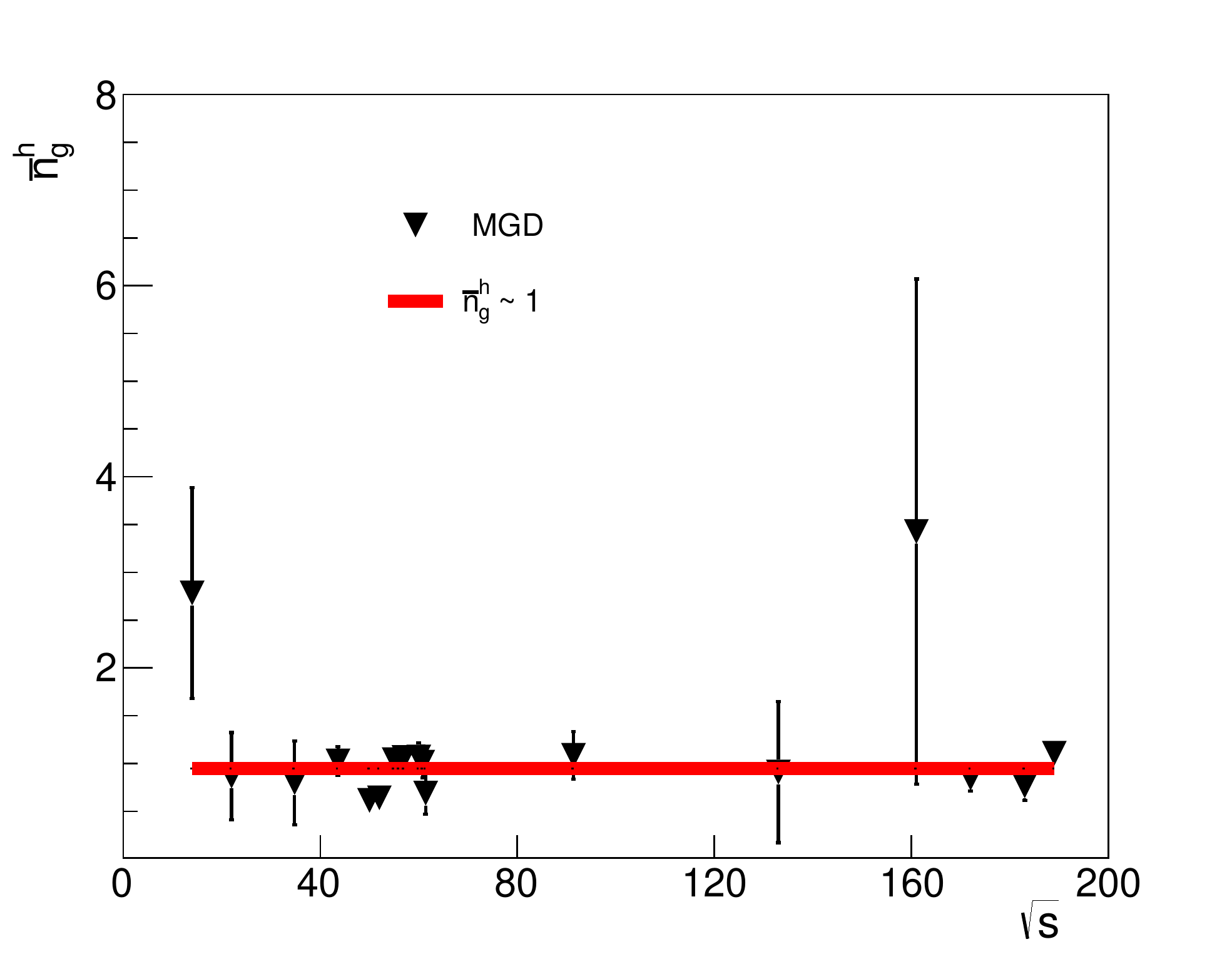}
}
\caption{Hadronization parameter of gluons $\overline n^h_g$ .}
\label{Fig:Fig1}
\end{figure}

The description of topological cross sections with taking into account the high multiplicity  region in the framework of the GDM is added in section 2. Section 3 demonstrates how the GDM improves this description with inclusion of a gluon fission. Estimation of the charge-exchange contribution in the GDM is carried out in section 4. Section 5 states the conclusions.

\section{Gluon dominance model}
The gluon dominance model (GDM) \cite{GDM1,GDM2} is a modification of the two-stage model (TSM) \cite{TSM, NPCS}. The TSM describes well multiplicity distributions in $e^+e^-$ annihilation in a wide energy region. It is based on the convolution of a QCD quark-gluon cascade  of an initial quark pair and a hadronization stage. In accordance with experimental data, the binomial distribution has been chosen for the description of hadronization. 

\begin{figure}[t]
\centerline{
\includegraphics[width=14cm]{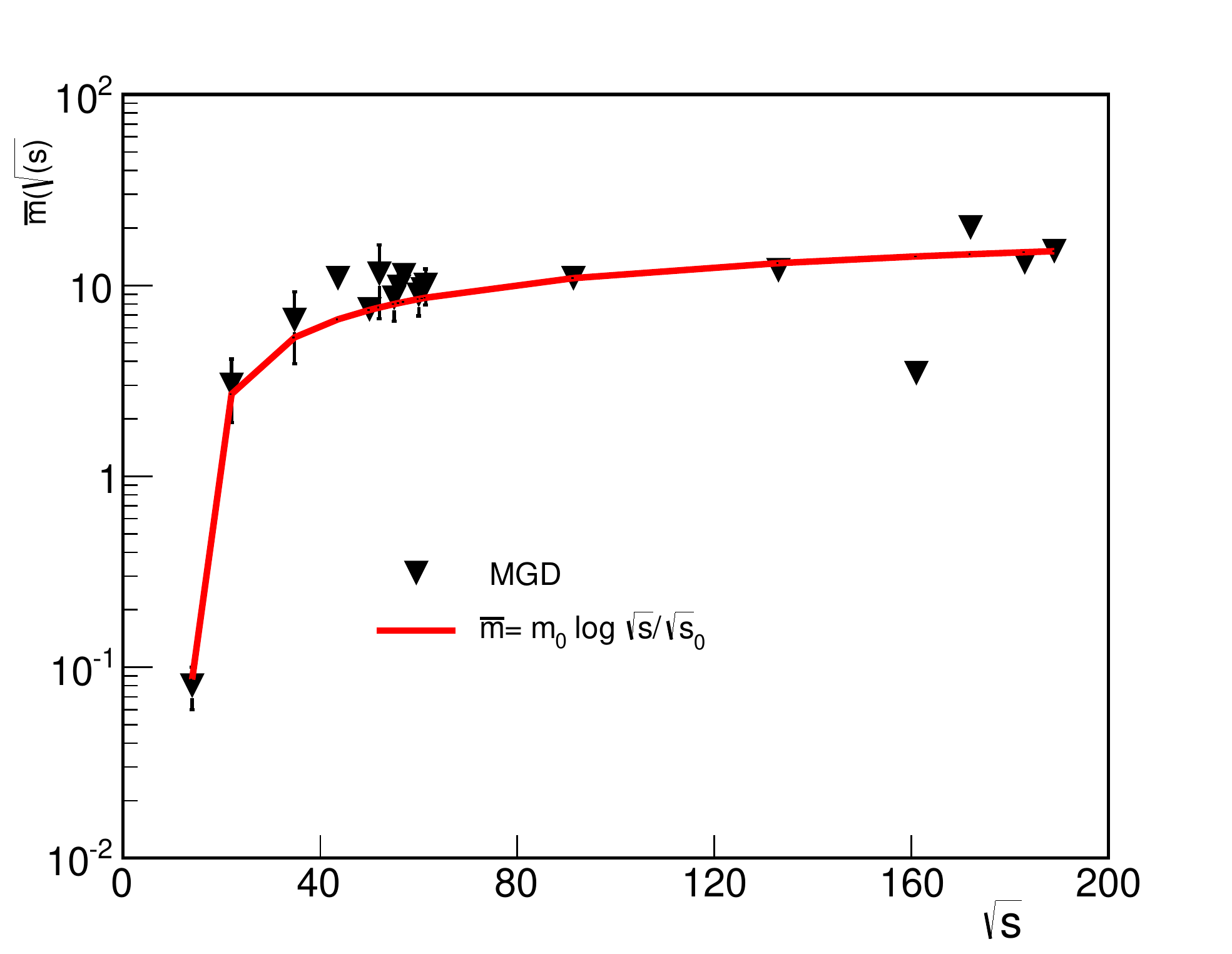}
}
\caption{Mean multiplicity of active gluons in TSM.}
\label{Fig:Fig2}
\end{figure}

The binomial distribution uses the following  parameters: the average multiplicity $\overline n^h_{q(g)}$ and the maximum number of hadrons $N_{q(g)}$ resulting from quark (gluon) in their passing through this stage. The ratio of the quark and gluon parameters, $\alpha = \overline n^h_{g}/\overline n^h_{q}$,  reduces the number of parameters at the hadronization stage to three. As it is shown in figure \ref{Fig:Fig1} the parameter  $\overline n^h_{g}$ is remaining approximately constant and close with 1 in a wide energy region (from 10 up to 200 GeV). That behaviour of the TSM's parameter is the confirmation of the fragmentation mechanism of hadronization in vacuum: one gluon fragments in one hadron. The main sources of secondary hadrons are gluons which are named active. Their mean multiplicity grows with energy and it can be approximated by the logarithmic dependence 
$\overline m~=~m_0 \log \sqrt s/\sqrt s_0$ shown in figure \ref{Fig:Fig2}, where $\sqrt s$ is the c.m.s. (center-of-mass system) energy and $m_0$ and $s_0$ are the parameters. 

As opposed to $e^+e^-$ annihilation in \textit{pp} interactions, it should take into account valency quarks and nascent gluons which form an initial quark-gluon system. The comparison with experimental data shows that multi-particle production is implemented by active gluons. Valency quarks are staying in the leading particles as passive spectators \cite{GDM1,GDM2}. Just in this case the description of topological cross sections becomes successful. Therefore, this model approach has got the name of the gluon dominance model (GDM). Two main schemes were considered: with and without gluon fission. The comparison with experimental data was carried out in both schemes which present convolution of two stages. 

The quark-gluon system is formed and it is developed owing to active gluons. These gluons can give a branch. This stage is described by the Poisson distribution with or without addition of the Farry distribution (accounting of gluon fission). The hadronization (second stage) is described by the binomial distribution. The topological cross sections in the scheme without gluon fission have the following form  

\begin{equation}
\sigma _n=\sigma _{inel}\sum\limits_{m}\frac{e^{-\overline m} \overline
m^{m}}{m!} {mN_g\choose n-2} \left(\frac{\overline n_g^h}
{N_g}\right)^{n-2}\left(1-\frac{\overline n_g^h} {N_g}\right)^{mN_g-n+2}
\label{eq.gluon1}
\end{equation} 

The GDM describes well all experimental topological cross sections for $pp$ interactions from U-70 energy up to ISR energies \cite{GDM2}. The hadronization parameter of gluons $\overline n^h_{g}$ grows from 1.5 at 50 GeV/c, U-70, up to 3.3 at 62.2 GeV, ISR \cite{GDM1,GDM2}. The growth of the hadronization parameter is the evidence of the  implementation of the recombination mechanism of hadronization. This stage occurs in quark-gluon medium but not in vacuum.
\begin{figure}[t]
\centerline{
\includegraphics[width=19 cm]{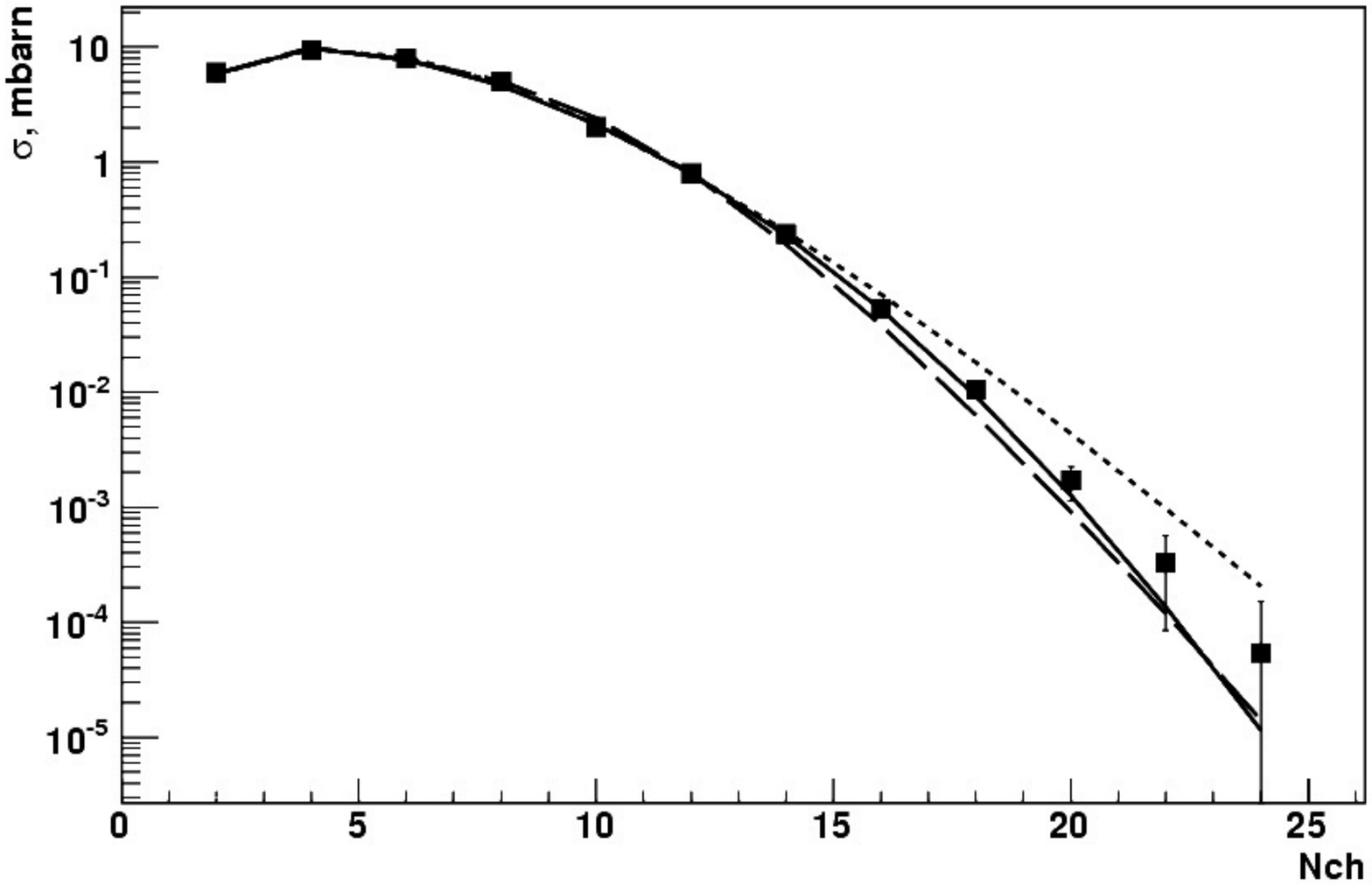}
}
\caption{Description of $\sigma (N_{ch})$ data \cite{SVD,SVD1} by GDM (\ref{eq.gluon1}) \cite{GDM1,GDM2} (solid line), model \cite{IHEP} (long dashed line) and the negative binomial distribution \cite{NBD} (short dashed line).}
\label{Fig:Fig3}
\end{figure}
In figure~\ref{Fig:Fig3} the topological cross sections at 50 GeV/c are presented. The region of low multiplicity was measured by the Mirabelle Collaboration \cite{Mirab}. The high multiplicity region was supplemented by the SVD Collaboration \cite{SVD,SVD1}. The GDM (\ref{eq.gluon1}) is presented in this figure by the solid line. The second model based on the KNO-function obtained from the relations between the elastic and inelastic cross sections \cite{IHEP} is shown by a long dashed line, and the third model based on the negative binomial distribution \cite{NBD} is denoted by a short dashed line. These models describe well the region of small multiplicity, but in the high multiplicity region small overestimation is being observed  for the negative binomial distribution \cite{SVD,SVD1}.
 
\section{Gluon fission} 
It is known that gluons give branching at high energies. It becomes predominating and gives an additional contribution to multiplicity.  In the GDM it is taken into account in the following way:
\begin{figure}[t]
\centerline{
\includegraphics[width=14cm]{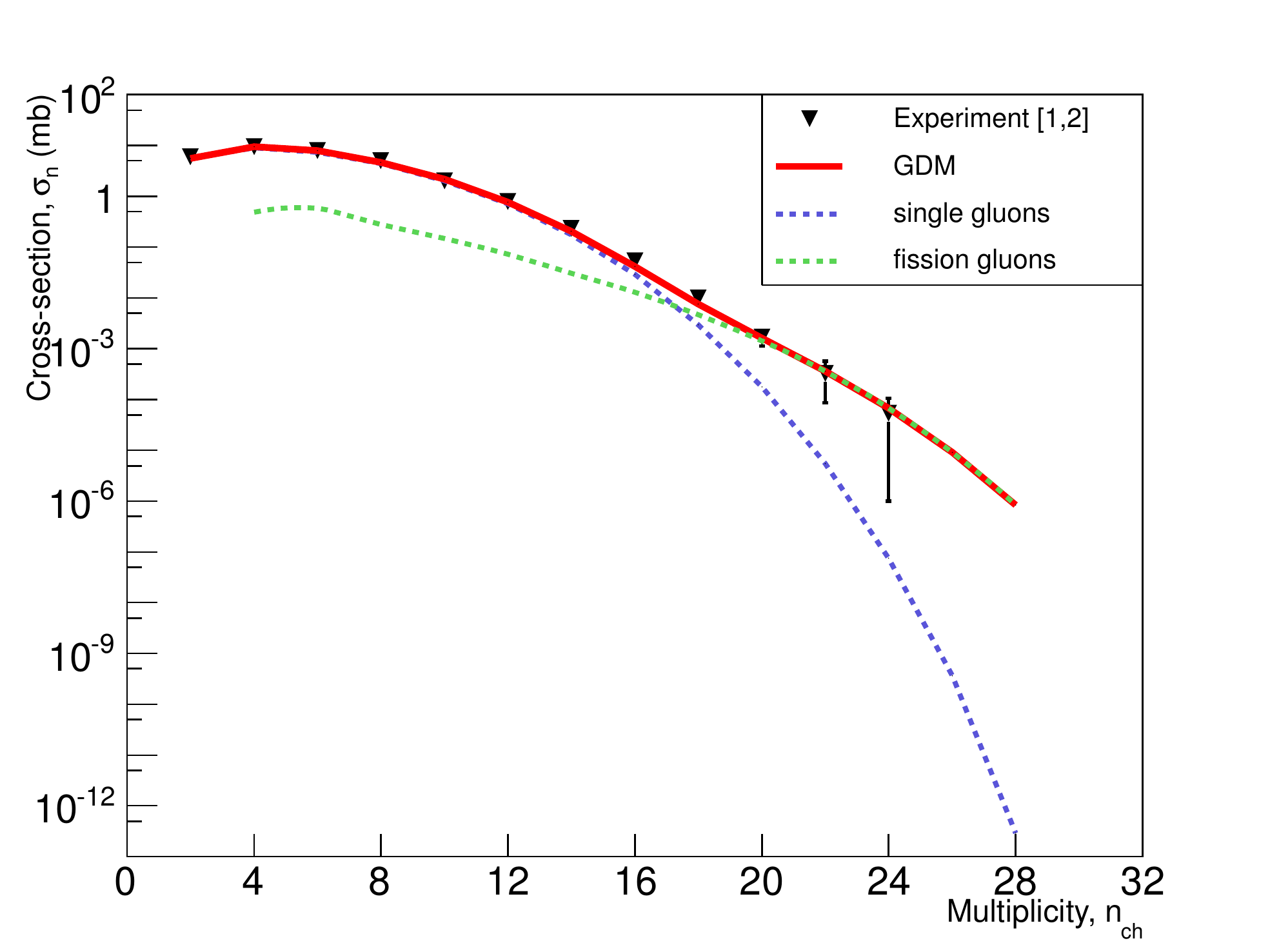}
}
\caption{Topological cross sections $\sigma _n$ versus charged multiplicity $n_{ch}$ in the GDM. The dashed blue line describes the contribution of single sources, the green line is the sources consisting of two gluons of fission, the solid red line is their sum.}
\label{Fig:Fig4}
\end{figure}
$$
\sigma _n=\alpha _1\sum\limits_{m_1}\frac{e^{-\overline m_1} \overline
m_1^{m_1}}{m_1!} {m_1N_g\choose n-2} \left(\frac{\overline n_g^h}
{N_g}\right)^{n-2}\left(1-\frac{\overline n_g^h} {N_g}\right)^{m_1N_g-n+2}+
$$
\begin{equation}
+\alpha _2\sum\limits_{m_2}\frac{e^{-\overline m_2} \overline
m_2^{m_2}}{m_2!} {2m_2N_g\choose n-2} \left(\frac{\overline n_g^h}
{N_g}\right)^{n-2}\left(1-\frac{\overline n_g^h} {N_g}\right)^{2m_2N_g-n+2},
\label{gluon2}
\end{equation}
where $m_{1,2}$ are multiplicities of single and double gluon sources, $\overline n_h^h$ and $N_g$ are the parameters of hadronization of gluons. In this scheme at the second stage the branching gluons give twice times as large hadrons  because they are  sources consisting of two gluons (second summand in $\sigma _n$). The ratio of two parameters $\alpha _1/\alpha _2$ is approximately equal to 1.8. The rest of the parameters do not almost change in comparison with (\ref{eq.gluon1}).

In the double-logarithmic approximation \cite{kur} it was revealed that the emission of two gluon jets formed by the fission process of a paternal quark (antiquark) is predominant with increasing energy and can explain the angle broadening of secondary particles. This fission can give the broadening of topological cross sections at high multiplicity \cite{GDM1,GDM2,kur}. In figure \ref{Fig:Fig3}, the comparison of measured topological cross sections \cite{Mirab,SVD} with the GDM (solid red line) is shown. In accordance with the GDM and taking into account the gluon fission the topological cross section consists of two summands. The dashed green line presents the contribution that appeared from the single gluon sources; the second summand (dashed blue line) in (4) describes the contribution responsible for fission gluons. Obviously, taking into account the gluon fission  slightly improves the description of data.

The GDM has been modified for describing the topological cross section in the proton-antiproton annihilation by introducing intermediate charged quark topologies. This modification is based on the following things: well-known  experimental fact of the leading of two charged pions and an active role of gluons in multi-particle production of hadrons, which follows from  the GDM. Quark topologies are determined by the formation of three neutral pions or two charged and one neutral pion from valent quarks only. Also, valent quarks can be combined with quarks from the quark-gluon system. In this case, the number of pions with valent quarks can increase up to 6. This scheme describes well the topological cross sections $\Delta \sigma _n$ of the pure $p\overline p$ annihilation process, which are  defined as  differences $\Delta \sigma _n =\sigma_n(p\overline p) -\sigma _n(pp)$ \cite{GDM1}.

\section{Charge-exchange in \textit{pp} interactions}
The experiments show that the charge-exchange can be realized at the scattering of protons off hydrogen or nucleus targets \cite{charg}. In this case, one of the protons gives its charge to a neutral pion and turns into a neutron

\begin{equation}
p + p \to p + \pi ^+ + n + N_{\pi }.
\label{eq:aa}
\end{equation}
In (\ref{eq:aa}) $N_{\pi}$  is the number of secondary pions and $n$ is a neutron. The estimation of the charge- exchange can be carried out in the inelastic channel 2 $\to $ 2.  In accordance with the Mirabelle data~ \cite{Mirab}, the elastic and inelastic cross sections are equal to $\sigma _{2,el}$ = 6.90 mb and $\sigma _{2,inel}$ = 5.71~mb, respectively. Their ratio is equal to $r =\sigma _{2,el}/\sigma _{2,inel}$ = 0.82. In the MGD the cross section is $\sigma _{2,el} \sim \exp ({-\overline m})$ because the active gluons do not appear. Since we  do not know a contribution of the charge-exchange, we can use for $\sigma _{2,inel}$ the expression $r\cdot \exp({-\overline m})$.
At the same time $\sigma _{2,inel}$  can be represented as a sum of two summands
\begin{equation}
\sigma _{2,inel} = \sigma ^{+ch}_{2,inel}  + \sigma ^{-ch}_{2,inel},
\end{equation}
with a charge-exchange (+ch) and without a charge-exchange (-ch). The GDM estimates the second summand in (6) the  the following way:
\begin{equation}
 \sigma ^{-ch}_{2,inel}\sim  \sum\limits_{m}\frac{e^{-\overline m} \overline
m^{m}}{m!}
\left(1-\frac{\overline n^h} {N_g}\right)^{m\cdot N_g-n+2},
\end{equation}
when active gluons appear without formation of charged particles. Let $P = \sigma _{2,inel}/\sigma ^{-ch}_{2,inel}$ , then instead  $\sigma _{2,inel}$ we use expression (7) with factor P and from fitting find $P$ = 2.18. Hence the coefficient of the charge-exchange
$$
q = \sigma ^{+ch}_{2,inel} /\sigma _{2,inel}\cdot 100\%
$$
consists of $\simeq $ 50 \% \cite{Egle}. This estimation corresponds to the experimental data \cite{charg}. The contribution of the charge-exchange is essential, and at small multiplicity it is necessary to have that in mind.

\section{Conclusion}
In this review, the main results of high multiplicity study within the GDM have been presented. The region of high multiplicity is unique, promising and fruitful. We expect that new collective phenomena will be
found out and our Collaboration will advance significantly in understanding the multi-particle production especially at the hadronization stage.

\section{Acknowledgment}
I thank all participants of the SVD-2 Collaboration for their active work and useful discussions.


\section*{References}
\begin{thebibliography}{100}
\bibitem{SVD} Ryadovikov VN 2012 Topological cross sections in proton-proton interactions at 50 GeV \emph{Phys.Atom.Nucl.}  \textbf{75}, Issue 3, 315-320.
\bibitem{SVD1} Kokoulina E~S, Nikitin V~A, Petukhov Yu~P and Kutov A~Ya 2012 Proton interactions with high multiplicity \emph{Phys.Atom.Nucl.} \textbf{75}, Issue 6, 664-667.
\bibitem{SVD2} Avdeichikov V~V \emph{et al} 2013 Spectrometer with a vertex detector for experiments at the IHEP accelerator \emph{Instrum.Exp.Tech.} \textbf{56} 9-31.                             
\bibitem{Trig} Avdeichikov V~V \emph{et al} 2011 A trigger of events with a high multiplicity of charged particles at the SVD-2 setup \emph{Instrum.Exp.Tech} \textbf{54} 159-168.                         
\bibitem{Phenom} Kokoulina E~S, Kutov A~Ya, Nikitin V~A and Popov V~V  2011 Analysis of high multiplicity events  \emph{Phys.Part.Nucl.Lett.} \textbf{8} 855-859.    http://dx.doi.org/
\bibitem{Phenom1} Kokoulina E S, Nikitin V A, Petukhov Yu P, Karpov A V and Kutov A Ya 2010  Search for collective phenomena in hadron interactions \emph{Phys.Atom.Nucl.}  \textbf{73} 2116-2124.
\bibitem{Mirab} Ammosov V~V \emph{et al} 1972 Phys.Lett. B42 (1972) 519-521.    http://            
\bibitem{Fluct} Kokoulina E S 2011 Neutral Pion Fluctuations in pp Collisions at 50 GeV by SVD-2 \emph{Prog.Theor.Phys.Suppl.} 193 306-309.
\bibitem{Meson} Afonin A G \emph{et al} 2012 Neutral pion number fluctuations at high multiplicity in pp-interactions at 50 GeV \emph{EPJ Web Conf.}  \textbf{37} 06002.   http://dx.doi.org/10.1051/epjconf/20123706002
\bibitem{FluRiad} Ryadovikov V N 2012 Fluctuations of the number of neutral pions at high multiplicity in pp interactions at 50 GeV \emph{Phys.Atom.Nucl.} \textbf{75}, 989-998.
\bibitem{ICHEP12} Kokoulina E for SVD-2 Collaboration 2013 Evidence for a pion condensate formation in pp interactions at U-70 \emph{PoS ICHEP2012} 259
\bibitem{Bald12} Kokoulina E S 2012 The evidence for the pion condensate formation in pp interactions at U-70 \emph{PoS Baldin-ISHEPP-XXI} 007
\bibitem{Land} Landau L D and Lifshitz E M 1969 Statistical Physics ( Volume \textbf{5} of A Course of Theoretical Physics ) Pergamon Press.
\bibitem{Goren1} Begun V V and Gorenstein M I 2007 Bose-Einstein condensation of pions in high multiplicity events. \emph{Phys. Lett.} \textbf{B} 653, N2-4, 190-195.
\bibitem{Goren} Begun V V and Gorenstein M I 2008
    Power Law in Micro-Canonical Ensemble with Scaling Volume Fluctuations \emph{Phys.Rev.}  \textbf{C}78 024904.
\bibitem{Ulery}  Ulery J G. STAR Collaboration. Are there Mach cones in heavy ion collisions? Three-particle correlations from STAR. 2005 \emph{Int.J.Mod.Phys.} \textbf{E}16 2005--10
\bibitem{Chlia} Chliapnikov P \emph{et al.} 1984 Observation of direct soft photon production in K+ p interactions at 70-GeV/c. \emph{Phys.Let.} 141\textbf{B} 276-280.
\bibitem{Braz}Kokoulina E S, Kutov A Ya and Nikitin V A 2007 Gluon dominance model and cluster production \emph{Braz.J.Phys.} \textbf{37} 785-787.
\bibitem{cher}  Kokoulina E S and Kutov A Ya 2008 High-multiplicity study \emph{Phys.Part.Nucl.Lett.}  \textbf{71}, No 9, 1573-1581.
\bibitem{GDM1} Kokoulina E S and Nikitin V A 2005 Study of multiparticle production by gluon dominance model \emph{17th International Baldin Seminar on High Energy Physics Problems: Relativistic Nuclear Physics and Quantum Chromodynamics (ISHEPP 2004)}
\bibitem{GDM2} Ermolov P F, Kokoulina E S, Kuraev E A, Kutov A Ya, Nikitin V A, Pankov A A, Roufanov I A and Zhidkov N K 2005 Study of multiparticle production by gluon dominance model. Part II. \emph{17th International Baldin Seminar on High Energy Physics Problems: Relativistic Nuclear Physics and Quantum Chromodynamics (ISHEPP 2004)}
\bibitem{SP} Volkov M K, Kokoulina E S and Kuraev E A 2004 Excitation of Physical Vacuum, \emph{Phys.Part.Nucl.Lett.}  Part 5, 16-23.
\bibitem{Barsh} Barshay S 1989 Anamolous soft photons from a coherent hadronic phase in high-energy collisions. \emph{Phys.Lett.} \textbf{B}227, N2, 279-284.
\bibitem{SP1} White Paper http://theor.jinr.ru/twiki-cgi/view/NICA/NICAWhitePaper.
\bibitem{TSM} Kokoulina E S  2002 Analysis of multiparticle dynamics in $e^+e^-$ - annihilation into hadrons by two stage model. XXXII ISMD, Alushta, Ukraine \textit{W.Sc. publ.}   340-343
\bibitem{NPCS} Kokoulina E S 2006 Gluon Dominance Model and $e^+e^-$ annihilation, Proceedings of the 13th annual seminar "Nonlinear Phenomena in Complex Systems", Minsk, May 16-19,  Joint Institute for Power and Nuclear Research-Sosny, NAS of Belarus \textbf{13} 73-82.
\bibitem{IHEP} Semenov S V, Troshin S M, Tyurin N E and Khrustalev O A 1975 Connection Between Elastic and Inelastic Processes at High-Energies. (In Russian) \emph{Yad. Fiz.} \textbf{22} 792-800.
\bibitem{NBD} Giovannini A 1979 QCD jets as Markov branching processes \emph{Nucl.Phys.} \textbf{B}161 429-448.
\bibitem{kur} Kuraev E A, Bakmaev S and Kokoulina E S 2011 Azimuthal correlation of gluon jets created in $pp$, $p\overline p$ and $e^+e^-$ collisions, \textit{Nucl.Phys.}  \textbf{B}851 551-564.
\bibitem{charg} Didenko L A, Murzin V S and Sarycheva L I 1979 Leading and charged exchange in pi-p interactions at 40 GeV (talk). Proceedings, 16th International Cosmic Ray Conference, Vol.6,  29-33.
\bibitem{Egle} Kuraev E A, Kokoulina E S and Tomasi-Gustafsson E 2015 Hard light meson production in (anti)proton-hadron collisions and charge-exchange reactions. \emph{Phys.Part.Nucl.Lett.} \textbf{12} , 1-7.

\end{thebibliography}
\end{document}